**Orbital transfer reveals material-selective structure in Rydberg excitons**

Mengli Wang[1†], Peng Chen[1†], Wenqi Xue[1†], Haonan Xia[1], Hong Zhang[2], Benniu Zhang[1*]

[1]Chongqing Jiaotong University, Chongqing 400074, China.
[2]State Key Laboratory of Mountain Bridge and Tunnel Engineering, Chongqing 400074, China.

*Corresponding author. Email address: benniuzhang@cqjtu.edu.cn
†These authors contributed equally to this work.

## Abstract

Rydberg excitons in atomically thin semiconductors offer a sensitive view of dielectric screening and electronic structure. As their spectra become increasingly well resolved, a central question is how much material information an effective Hamiltonian can predict beyond the states used to determine it. Using the 1s, 2p and 2s levels to determine a scalar Rytova–Keldysh Hamiltonian, we test its transfer to the excluded 3p state. The transfer deviations have opposite signs: the observed 3p lies 7.38 meV below the prediction in $MoSe_2$ and 5.58 meV above it in $WS_2$; the opposite signs persist across the complete reported ±2σ peak-error boxes. In a separate high-field-calibrated benchmark, RK removes 66.8% of the orbital error of a pure two-dimensional Coulomb reference, making the remaining state-selective deviations particularly informative. After separating gap and common-scale contributions, the residuals reveal distinct material-dependent orbital patterns and constrain simple multichannel descriptions. For $WS_2$, the measured 2p/3p curvatures yield a parameter-fixed prediction, $Q_{4p}$=+4.08 meV, for a future same-platform measurement. Together these results show how a well-performing spectral model can be tested for transferable physical information, with the next unmeasured state providing an independent experimental check.

## Introduction

Atomically thin semiconductors offer an unusual view of the interaction between light and matter [1]. A photoexcited electron and the hole it leaves behind can bind into an exciton, whose excited states form a Rydberg series reminiscent of hydrogen [2,3]. The number counts the shell and the letter the orbital shape. In a monolayer, the attraction is strongly shaped by the surrounding dielectric environment [4,5]. As the electron–hole separation changes from one Rydberg state to the next, their energies and sizes sample that interaction at different length scales [6–8]. The resulting ladder is becoming a detailed optical probe of screening, exciton size and electronic structure [9]. It also provides a way to assess dielectric engineering of two-dimensional heterostructures, materials of growing interest for optoelectronics and valley-based quantum technologies [4,10].

That opportunity also contains a simple but consequential inference problem. The Rytova–Keldysh (RK) interaction offers a compact, widely used account of nonlocal dielectric screening and reproduces many features of nonhydrogenic exciton spectra [11–13]. Indeed, Takahashi et al. reported that calculations using the RK potential reproduced the measured s- and p-state energies to within a few meV [14]. That agreement establishes descriptive accuracy, but it does not test whether a model calibrated on a specified subset transfers to a state excluded from parameter determination. Its parameters are commonly established from the same spectral ladder used to assess the result. A close fit therefore leaves an important question open: which features of the material has the model learned, and which can it predict in a different quantum state? Answering that question would help distinguish dielectric information carried by the entire ladder from the extra structure visible in selected orbitals.

To test this transfer question, we use same-location sum-frequency-generation spectra that resolve both s and p exciton states in $MoSe_2$ and $WS_2$, together with $WSe_2$ spectra from an earlier measurement by the same group [14,15]. We use the 1s and 2s energies to anchor the common gap and energy scale and the 2p state to determine the RK spectral shape. The 3p state then tests transfer beyond the calibration set with no additional parameter adjustment.

Here we develop a transfer-based spectral test that separates scalar screening from orbital-selective structure, identifies opposite material responses, constrains minimal multichannel descriptions, and produces a parameter-fixed prediction for the next $WS_2$ p state. We first establish the material dependence of the 3p deviation and then ask how much of the broader orbital spectrum scalar screening already describes. We next examine the material-dependent structure that remains after removing the common gap and energy scale. Finally, the measured $WS_2$ ladder determines a specific 4p target for future spectroscopy.

## Materials and methods

### Study design

This work is a theoretical and computational reanalysis of published exciton spectra. Primary transition energies and reported errors come from Takahashi et al. and the associated earlier $WSe_2$ work [14,15]; the underlying numerical spectra are deposited in the Kyoto University repository [20]. The calibration–transfer analysis uses their tabulated peak centers, with the 3p state excluded from model parameter determination.

### RK model and numerical prediction curve

The scalar RK potential is solved through the dimensionless central-potential radial eigenproblem with $q=r^*/a$. We compute an 800-point transfer manifold spanning q=0.25-15 using a cell-centered finite-volume solver; maximum dimensionless drift is below $1.45\times10^{-4}$ across the tested grid variants. The same solver confirms monotonicity of the required transfer ratios over this interval. Details appear in Supplementary Notes 1, 14 and 15.

### 3p prediction and uncertainty propagation

For each material the first three same-location peaks determine $S_{2p}$ and the scalar RK q shape; 3p then supplies the transfer residual $E_{3p}(\mathrm{obs})-E_{3p}(\mathrm{RK})$. The source peak centers are: $MoSe_2$ 1s/2p/2s/3p = 1.6390/1.7797/1.7983/1.8182 eV; $WS_2$ = 2.0601/2.1944/2.2033/ 2.2272 eV; $WSe_2$ = 1.6982/1.8203/1.8273/1.8453 eV [14,15]. Reported peak errors are 0.6 meV for $MoSe_2$,

0.6/0.6/0.6/0.7 meV for $WS_2$ and $WSe_2$. These reported errors combine spectral resolution (~0.6 meV after the 2025 source correction) and fitting error [21].

The published peak errors are propagated through deterministic 16-corner calculations at ±1σ, ±2σ and ±3σ and through 200,000 independent-Gaussian draws using the stated settings. The corners describe sensitivity to simultaneous peak shifts; the Gaussian draws assess that particular error model rather than reconstruct the original experimental likelihood. A finite-volume implementation confirms the central residual signs and the ±2σ/±3σ sign boundaries; its $MoSe_2$ ±2σ limits differ from the manifold-interpolated bounds by approximately 0.2 meV, with agreement on sign and all qualitative conclusions (Supplementary Note 15).

**One-parameter extensions and benchmark models**

The p-sector extensions, reciprocal transfer, pure-Coulomb comparison and $WSe_2$ high-n separation use the separate high-field-q calibration, together with each sample's 1s–2s energy scale (Supplementary Note 3). The 66.8% reduction is the held-out 2p/3p benchmark under that protocol, with pure Coulomb serving as an illustrative lower-complexity reference. The reverse p-specific RMSE describes the numerical fit under that protocol.

**Independent size and screening checks**

Secondary exciton-size controls draw on published $MoSe_2$ and $WSe_2$ studies [2,7,8]. Selected zero-field $MoSe_2$ peak positions, where the precise numerical centers were not available in the materials used for this comparison, were digitized from a published figure and checked under ±2-meV reading perturbations (Supplementary Note 5). These auxiliary values enter neither the primary same-location 3p calibration nor the $WS_2$ 4p target. Separate published dielectric-tuning studies provide context for the screening coordinate [9].

**Fixed-*q* curvature and single-pole self-energy model**

At an externally fixed q, the quotient $Q_p$ removes additive-gap and common-scale directions from the observed-minus-RK residuals. The resulting divided differences are examined within a passive, shared scalar positive-measure single-pole construction. The inference is conditional on the chosen RK reference, common-residue assumption and projected self-energy. A central positive-weight solution exists for $WS_2$, while $MoSe_2$ and $WSe_2$ lie outside the specified minimum construction in different ways (Supplementary Notes 6 and 7).

Exploratory self-consistent Hermitian calculations (Supplementary Note 7A) test the identity of the observed branches and construct positive-weight multipole alternatives from the existing measured levels. Their RK-like projector weights are model coordinates rather than measured optical intensities. The 4p continuations demonstrate why additional information on symmetry-allowed dark-state energies and couplings is needed for robust exclusion ranges in $MoSe_2$ and $WSe_2$.

**Berry-curvature sensitivity scale**

A three-band nearest-neighbor exciton-geometric calculation [22,23] yields a $MoSe_2$ 2p $m = \pm 1$ splitting of 15.9 meV, used in the Discussion as a sensitivity scale for geometric orbital effects. Input parameters and derived quantities are listed in Tables S2 and S3.

## Results

A frozen screening model reveals opposite-sign 3p deviations

The first test is deliberately simple. Once a gap and overall energy scale are factored out, scalar RK screening has only one dimensionless shape coordinate:

$$E_{\mathrm{n}}^{\mathrm{scr}} = E_g + S\varepsilon_{\mathrm{n}}(q)$$

The offset $E_{\mathrm{g}}$ translates the ladder, $S$ sets its energy scale and q determines the relative level spacings. Two ratios capture the calibration and the subsequent transfer:

$$\mathrm{S}_{2\mathrm{p}} = \left(E_{2\mathrm{p}} - E_{1\mathrm{s}}\right)/\left(E_{2\mathrm{s}} - E_{1\mathrm{s}}\right); \mathrm{T}_{3\mathrm{p}} = \left(E_{3\mathrm{p}} - E_{2\mathrm{s}}\right)/\left(E_{2\mathrm{s}} - E_{1\mathrm{s}}\right)$$

The 1s, 2p and 2s peaks determine $S_{2p}$ and thereby select a unique q on the computed RK manifold over q=0.25-15. With that choice, the ratio $T_{3p}$ is fixed before the 3p peak enters the comparison. This split has a clear spectroscopic purpose: the first p state participates in calibration, while the next p state probes how faithfully the description transfers along the same orbital series.

The outcome is a striking difference between materials. The 3p residual (observed minus predicted) is -7.38 meV in $MoSe_2$, +5.58 meV in $WS_2$ and +2.22 meV in $WSe_2$. (Fig. 1). For the $MoSe_2/WS_2$ pair, every corner of the complete reported ±2σ four-peak error boxes retains the opposite signs: the ranges are -12.31 to -2.55 and +1.00 to +10.10 meV, respectively. The smaller $WSe_2$ shift provides a useful third comparison, although its uncertainty gives it less weight in the sign argument. The error model, broader ±3σ tests and alternative calibration splits are reported in Supplementary Notes 2 and 14.

A separate room-temperature, plasmon-enhanced $WS_2$ experiment provides an independent same-sign comparison under the same extrapolation protocol, despite its very different temperature, dielectric environment and optical architecture (16; Supplementary Note 12).

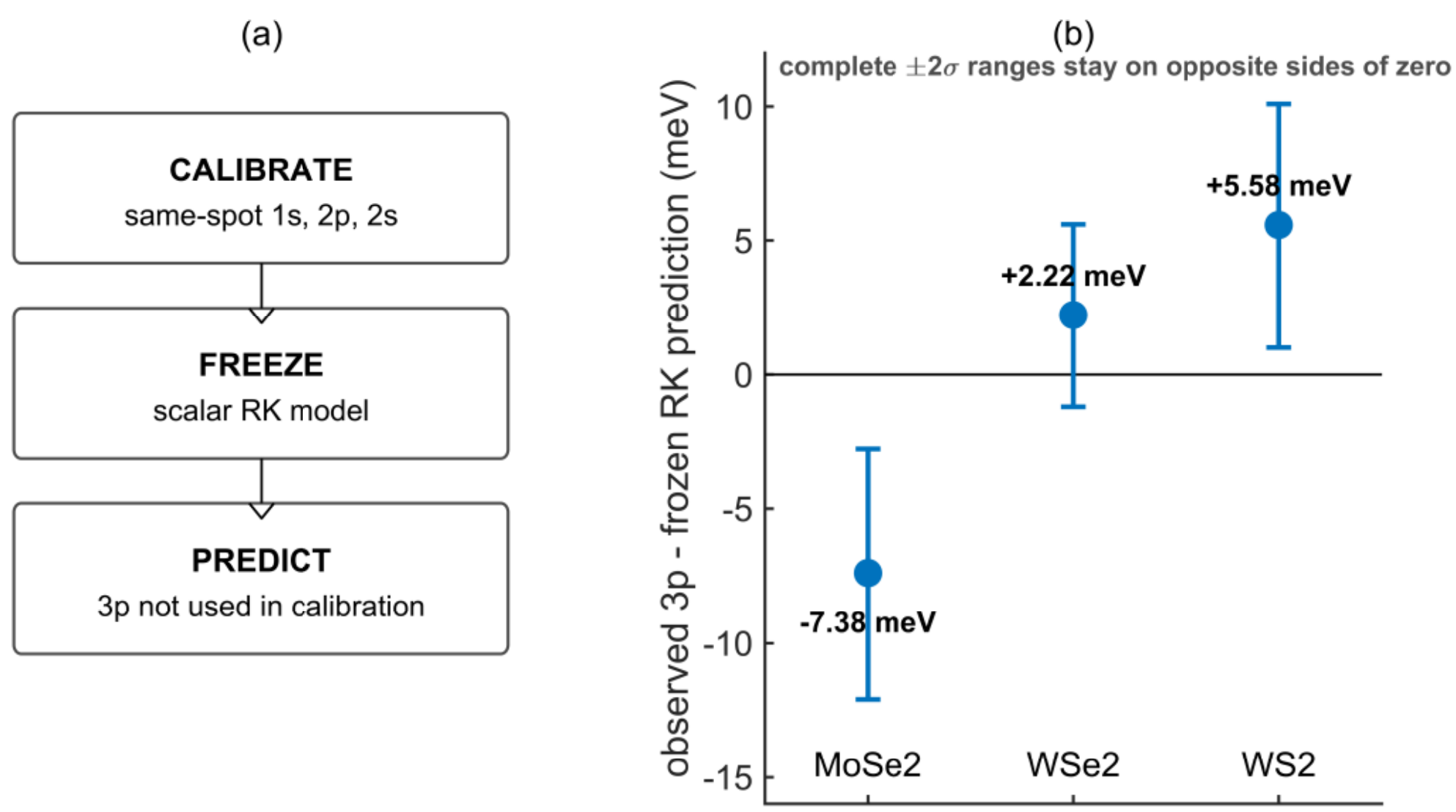

**Fig. 1. A held-out orbital reveals a material-dependent deviation.** (a) The 1s, 2p and 2s energies determine the scalar RK model, which is then held fixed for the 3p test. (b) The 3p residual (observed minus predicted) is −7.38 meV in $MoSe_2$, +2.22 meV in $WSe_2$ and +5.58 meV in $WS_2$. Reported ±2σ peak-error boxes preserve the opposite $MoSe_2/WS_2$ signs. The boxes show sensitivity to simultaneous peak perturbations rather than confidence intervals; the extreme ±3σ $WS_2$ box reaches across zero (Fig. S4).

The obvious next question is whether this contrast simply reflects an inadequate screening baseline. If scalar RK were too crude to capture the ladder in the first place, the 3p remainder would carry little additional information. We therefore ask how much of the broader orbital spectrum RK already explains before interpreting what it leaves behind.

**Screening captures the ladder while individual orbitals retain detail**

For this comparison we use high-field literature parameters to fix the RK screening coordinate and the measured 1s–2s splitting of each sample to restore the energy scale [2,6]. Both p states are excluded from the scalar calibration. Three simple p-sector extensions are then calibrated on 2p and evaluated on 3p: a separate p-specific RK coordinate, a constant p-channel defect and a localized $N^{-3}$ correction. Across the three materials, the scalar 3p RMSE of 4.34 meV becomes 4.18, 4.10 and 4.19 meV, respectively. These modest changes show how much of the discrepancy survives three economical ways of adding orbital freedom. More general screening and multichannel interactions remain open physical possibilities.

The direction of transfer also matters. A p-specific parameter determined from 3p and carried back to 2p produces approximately 18 meV RMSE, compared with approximately 2.9 meV for scalar RK in that reverse comparison. The asymmetry makes this simple additional coordinate a poor transferable correction across the measured p states. Supplementary Note 3 examines the parameter-search landscape behind this asymmetry.

A second result places these errors in perspective. Under the same high-field-q calibration, the aggregate held-out 2p/3p RMSE falls from 11.10 meV for a pure two-dimensional Coulomb reference to 3.68 meV for scalar RK: a 66.8% reduction (Fig. 2B). This simple reference isolates the contribution of dielectric screening; its 2s–2p degeneracy prevents a matched primary three-state calibration. Exciton-size measurements in $MoSe_2$ and $WSe_2$ and pressure-dependent Rydberg spectroscopy provide complementary evidence that screening is an important physical coordinate [2,7,9]. The puzzle is therefore sharper: scalar RK captures much of the global spectrum, yet a smaller, organized orbital remainder survives when the model is transferred across states.

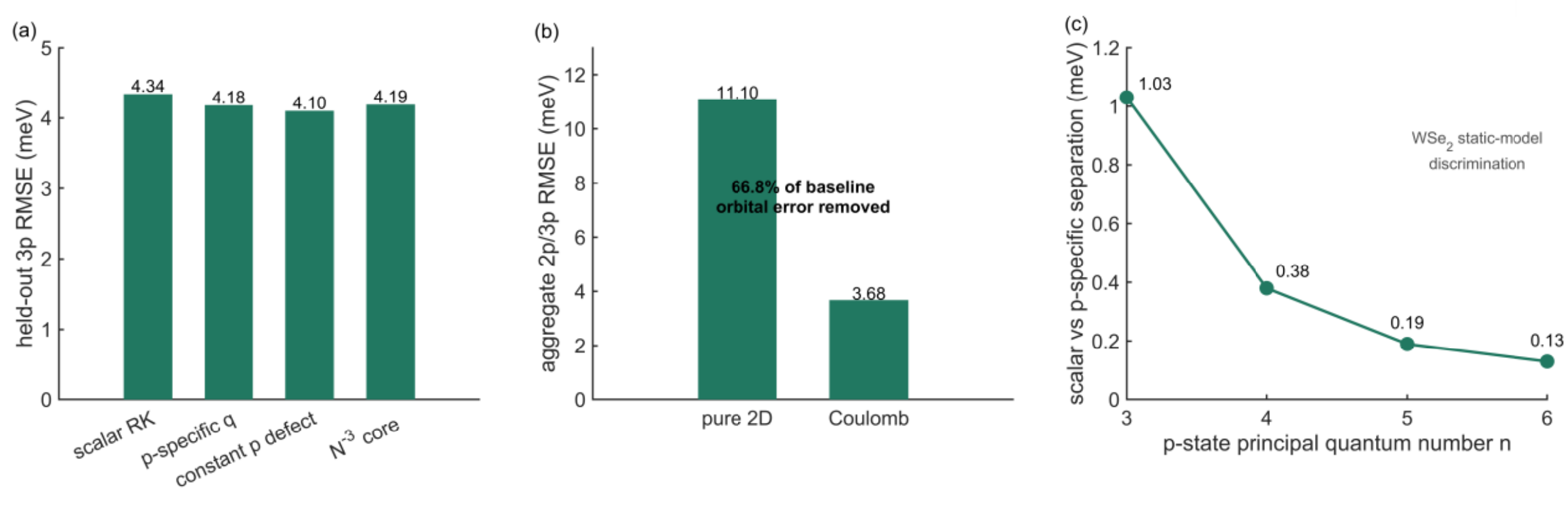

**Fig. 2. A useful screening model retains a state-selective remainder.** (a) With literature high-field q and same-sample 1s–2s scaling, three one-parameter p-sector extensions calibrated on 2p offer modest improvement when tested on 3p. (b) In that separate benchmark, scalar RK reduces the held-out orbital RMSE from 11.10 to 3.68 meV relative to pure 2D Coulomb, a 66.8% reduction. (c) For the stated $WSe_2$ p-specific comparison, the model separation falls from 1.03 meV at 3p to 0.13 meV at 6p. These results use a separate calibration from the primary Fig. 1 test.

This raises a natural question: if higher Rydberg states become increasingly hydrogen-like, why can state-specific information remain experimentally relevant? A key part of the answer lies in the different scales governing total binding and nearby-level separations.

**A more hydrogenic spectrum can still carry state-specific information**

For a two-dimensional Coulomb tail, the binding energy falls asymptotically as $n^{-2}$, whereas the separation of neighboring high-Rydberg levels falls as $n^{-3}$. A sufficiently localized correction in a penetrative fixed-angular-momentum channel can also have $n^{-3}$ diagonal and nearby-state matrix elements (17–19; Supplementary Note 4). Its contribution to total binding then becomes progressively smaller while remaining appreciable relative to the local level spacing. This scale separation offers a general physical way to understand how a nearly hydrogenic ladder can preserve sensitivity to additional structure. The measured 2p and 3p states lie below the asymptotic regime, so the argument is used here to interpret the information scales rather than to assign the source of their residuals.

The numerical comparison in $WSe_2$ makes the experimental consequence concrete. With q=1.680 fixed by the high-field analysis and one p-specific coordinate determined from 2p, the extra coordinate shifts the predicted 3p energy by 1.03 meV. Its effect falls to 0.38, 0.19 and 0.13 meV at 4p, 5p and 6p (Fig. 2C). For this particular contrast, resolving more distant members of the ladder would therefore require increasingly fine energy precision. The most informative measurement is the one that separates competing descriptions, rather than simply the state with the highest principal quantum number.

The remaining material dependence could still arise from familiar global differences such as the band gap or the overall binding scale. Factoring out those directions allows us to ask what orbital structure remains for comparison across materials.

**Removing global shifts exposes different orbital signatures**

For this comparison, we take independent high-field RK parameter sets as fixed reference shapes: q=2.665 for $MoSe_2$, 1.188 for $WS_2$ and 1.680 for $WSe_2$. Each sample's measured 1s–2s splitting supplies its energy scale [2,6]. The external coordinates are themselves inferred within RK analyses of other experiments; their values differ from those obtained by fitting the present same-location 1s/2p/2s triplets (Supplementary Note 14). The resulting comparison therefore asks a specific question: which orbital differences remain once gap and common-scale freedom are removed at those independently chosen reference shapes?

The quotient $Q_p$ (upper case, distinct from the screening coordinate q) removes an additive gap shift and a common rescaling exactly for a fixed RK shape. Differences between neighbouring shells then leave only the curvature. The resulting divided differences $C_{12;p}$ form three distinct pairs: (+0.625, −1.265) $eV^{-1}$ for $MoSe_2$, (−1.234, +1.341) $eV^{-1}$ for $WS_2$ and (−3.276, −0.376) $eV^{-1}$ for $WSe_2$ (Fig. 3). $MoSe_2$ and $WS_2$ retain opposite curvature patterns; $WSe_2$ traces a third direction. These signatures describe the measured residuals relative to the declared reference and can guide a more resolved examination of possible couplings.

These material-dependent curvature patterns can in turn constrain simple hidden-channel descriptions. To compare the three materials on common footing, we begin with the smallest passive representation that can reproduce the measured p-state curvature pair.

**What additional structure can the present spectra constrain?**
Within a fixed-q, passive common-scalar single-pole representation, $WS_2$ admits a positive spectral weight. The $MoSe_2$ curvature pair requires a negative weight at its central values and falls outside that minimum passive form. $WSe_2$ is incompatible with the particular projected shared-residue construction used for Fig. 3. For $WSe_2$, an alternative self-consistent one-pole realization becomes possible under a different branch assignment, making branch identity an important additional observable (Supplementary Note 7A).

These results delineate the information carried by the current level positions. Within the stated representation they distinguish a simple admissible $WS_2$ description from more constrained $MoSe_2$ and $WSe_2$ cases. Positive-weight multipole constructions can nevertheless reproduce all existing calibration energies while changing the predicted 4p shift. The current spectra therefore identify conditional model constraints and the kinds of independent measurements that could turn them into more specific microscopic assignments.

Additional dielectric, band-geometric and interseries effects remain natural subjects for those measurements. They include nonlocal or dynamical screening, intervalley mixing, nearby exciton channels and phonon-assisted processes. The present orbital fingerprints offer a way to organize such comparisons while keeping their microscopic interpretation open.

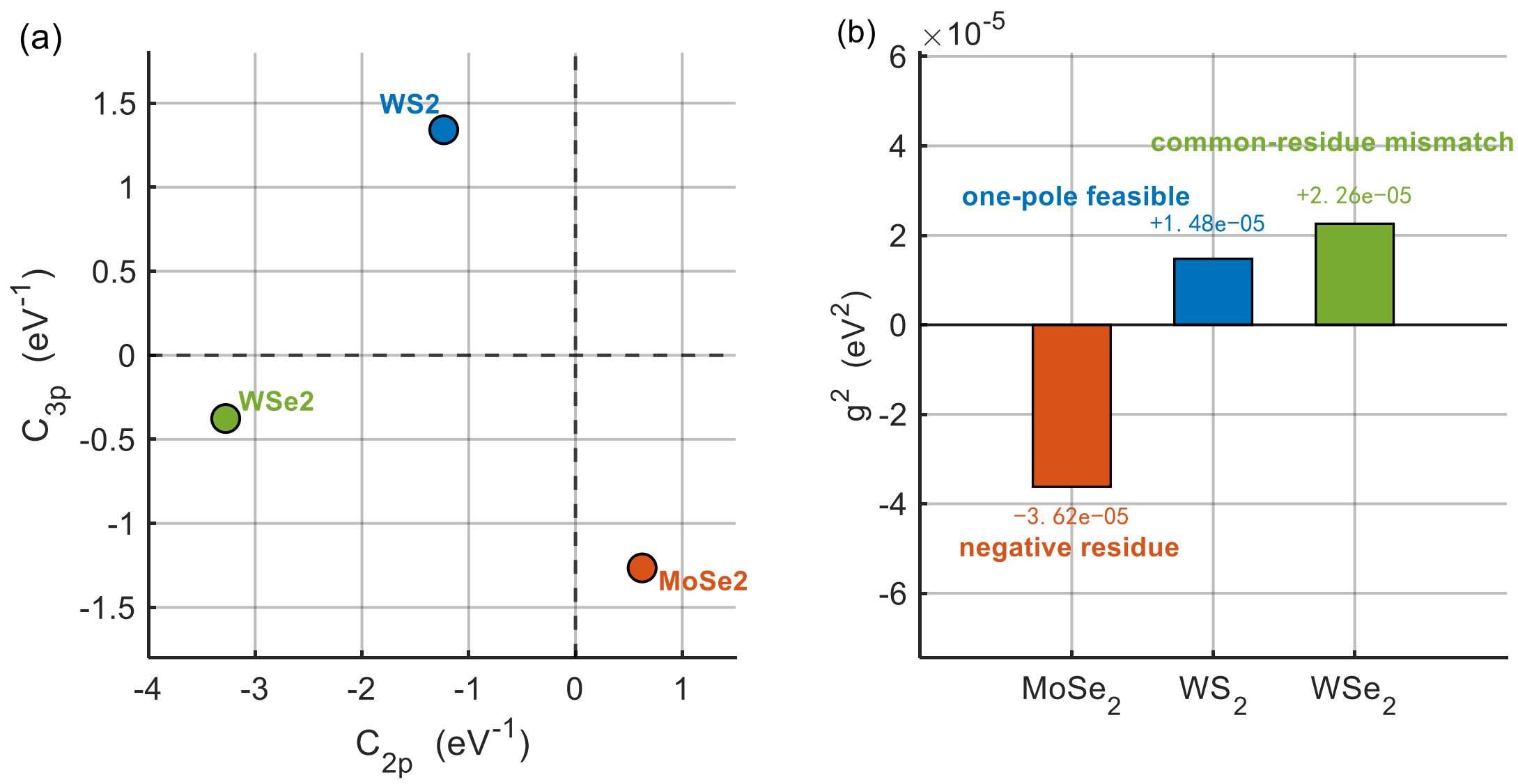

**Fig. 3. Material-dependent orbital curvature at a fixed screening reference.** (a) Removing gap and common-scale shifts from the 2p/3p energies leaves distinct divided-difference pairs in $MoSe_2$, $WS_2$ and $WSe_2$. (b) In the specified passive common-scalar single-pole projection, $WS_2$ has a positive-weight solution, $MoSe_2$ fails that minimum test at the measured central values and $WSe_2$ lies outside the tested shared-residue form. The classification is conditional on this representation; the alternative self-consistent $WSe_2$ branch and multipole examples are treated in Supplementary Note 7A.

**A frozen prediction for the next $WS_2$ orbital**
$WS_2$ then turns this residual structure into a prospective experimental test. The $WS_2$ curvature pair admits the selected positive single-pole continuation. Once its pole and spectral weight are determined by the existing 2p and 3p data at the fixed external q, the representation has no remaining scalar freedom with which to adjust the next p state. It therefore yields a concrete target for a future same-platform 4p measurement:

$$Q_{4p}\left(WS_2\right)=+4.08\ \text{meV}$$

The point target, +4.08 meV, remains frozen. Fixed-q numerical propagation gives a median near +4.0 meV and a conditional 68% interval of about +2.9 to +5.0 meV (Supplementary Note 13). That interval reflects the specified peak-error treatment within the selected model and will need to be supplemented by external-q and future experimental uncertainties when the new spectrum becomes available. The model family was selected using the measured 3p values, making the as-yet unmeasured 4p state the independent prospective test of this continuation.

The other two materials illuminate a complementary aspect of prediction. Present data can rule out a particular minimum construction. They cannot fix a unique next-state energy. $MoSe_2$ excludes the central positive-weight single-pole interpolation, whereas the $WSe_2$ single-pole alternative calls for a predominantly dark-state-like observed 2p branch in the tested basis. The multiple allowed higher-pole continuations make independent information on dark-state symmetry, energy and coupling particularly valuable before numerical 4p exclusion ranges can be established.

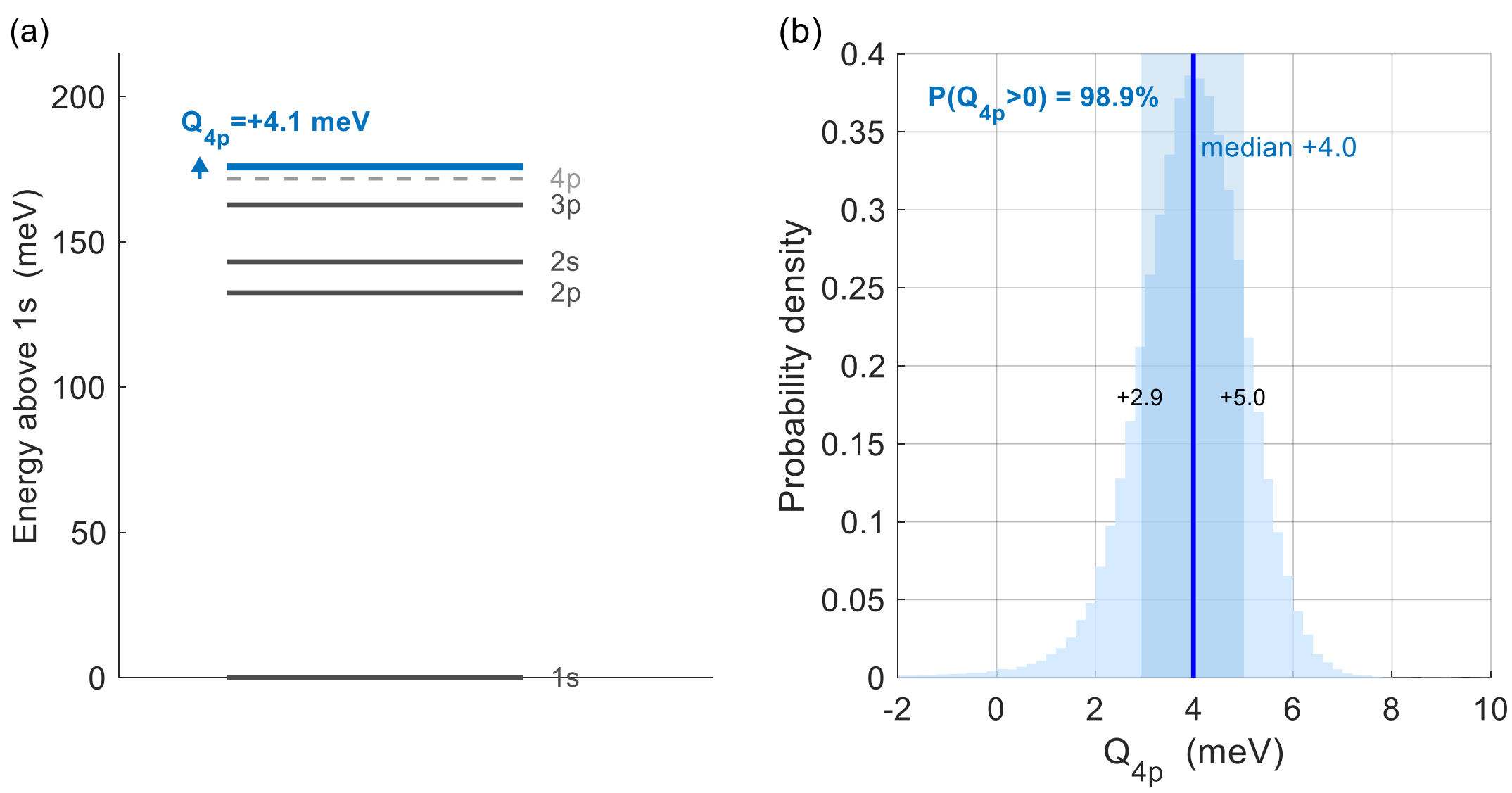

**Fig. 4. A frozen model sets a prospective $WS_2$ 4p target.** (a) The existing $WS_2$ 2p/3p curvatures fix a positive single-pole continuation, giving $Q_{4p}$ = +4.08 meV. (b) Fixed-q numerical propagation yields a median near +4.0 meV and a conditional 68% interval near +2.9 to +5.0 meV (Supplementary Note 13). The probability $P(Q_{4p} > 0)$ is evaluated under the fixed-q one-pole model with Gaussian peak-error propagation (200,000 draws); cross-device q variation and future experimental covariance are not included. A new 4p measurement on the same experimental platform can evaluate this previously frozen output once the independent parameter and measurement uncertainties are included.

## Discussion and conclusion

The central result is a separation between two kinds of success: reproducing a spectrum and carrying its physics forward to another state. In $MoSe_2$ and $WS_2$, a scalar screening model calibrated on three measured levels leaves 3p deviations of opposite sign. Independent benchmarks show how much the simple model already achieves: scalar RK removes 66.8% of the pure two-dimensional Coulomb orbital error, so the surviving 3p residuals reveal state-selective information on top of a screening model that already captures much of the spectrum. The residuals offer a more discriminating view of material dependence than the overall spectral fit alone.

More broadly, fitting accuracy and predictive transfer probe different aspects of an effective Hamiltonian: a compact model may describe complex spectra well, while carefully chosen excluded observables reveal how far that description carries beyond the states used to determine it. For two-dimensional semiconductors this distinction improves the interpretation of optical ladders used to infer screening and electronic structure, and it helps prioritize measurements that add independent physical information. The same approach offers a useful standard for other situations in which material properties are inferred from models fitted to increasingly rich spectra.

The wider set of sensitivity tests clarifies where that information travels. A second three-state split retains the $MoSe_2/WS_2$ sign pair, while other splits and independently inferred high-field screening coordinates expose the dependence on model calibration and experimental environment. The separate room-temperature $WS_2$ spectrum supplies a valuable same-sign comparison and, through its larger magnitude, illustrates the influence of sample architecture and measurement conditions. These features make repeated same-platform and independently prepared low-n spectra especially useful for establishing the stability of material-dependent transfer.

The broader physical picture follows the same distinction. Because total Rydberg binding and nearby-level spacings approach the Coulomb limit on different scales, increasingly hydrogenic energies can remain sensitive to finer state-dependent interactions. The three fixed-q curvature pairs exhibit just such residual diversity at the level of the measured energies. Passive dark-state tests then describe which simple mathematical continuations are consistent with those pairs and show how additional states or optical observables could distinguish among admissible microscopic realizations.

That distinction suggests two targeted experimental directions. In $WSe_2$, same-device polarization- and phase-sensitive spectroscopy, interpreted with independently measured transition matrix elements, could clarify the identity of the observed 2p branch. In $MoSe_2$, locating symmetry-allowed dark p-like channels and measuring their couplings would narrow the range of viable multichannel continuations. These experiments address different pieces of the inverse problem: one identifies the optical branch, while the other constrains the spectrum of the dark-state sector.

$WS_2$ already supplies the next decisive energy test. The positive single-pole continuation fixed by its measured 2p and 3p data gives $Q_{4p}$=+4.08 meV for a future same-platform measurement. A recorded 4p level, analysed with independent measurement errors and screening-parameter uncertainty, will directly assess that selected model. Its value lies in the fact that the model must now face a new datum using the parameters it already has.

**Conflict of interest**
**The authors declare that they have no conflict of interest.**

## Acknowledgments

**Funding:** B.N.Z. acknowledges the support by the National Natural Science Foundation of China (No. 52178272 and No. 52478300).

## Author contributions

M.W., P.C., and W.X. contributed equally to this work and share first authorship. They jointly conceived and designed the study, performed the theoretical and numerical analyses, interpreted the data, and drafted the manuscript. H.X. contributed to investigation and validation. H.Z. contributed to Supervision. B.Z. contributed to conceptualization, supervision, project administration, and writing—review and editing. All authors reviewed and approved the final manuscript.

**Data availability:** The primary experimental peak centers and their published errors come from Refs. [14,15,21]. All calculation code and generated data files are available as the Full-text calculation code and results package, which will be deposited in a public repository upon publication.

**AI-assisted preparation:** OpenAI ChatGPT and Doubao were used for manuscript organization, language editing, literature /provenance auditing, code review, and submission-package preparation. AI output was not treated as a source of scientific evidence. All scientific claims, source citations, numerical outputs, code, and final wording were reviewed by the authors, who retain full responsibility for the work.

# Supplementary Materials for

## Orbital transfer reveals material-selective structure in Rydberg excitons

Mengli Wang *et al.*

*Corresponding author: Benniu Zhang, benniuzhang@cqjtu.edu.cn

**Supplementary Materials provide the numerical definitions, reproducibility checks and model assumptions supporting the results in the main text:**

**Supplementary Note 1. RK prediction table, monotonic inversion, and numerical convergence**

The direct RK solution uses a single coordinate convention throughout. Let $C = e^2/(4\pi\varepsilon_0\kappa)$, $a = \hbar^2/(2\mu C)$, $r^* = r_0/\kappa$, $q = r^*/a$, and $E_1 = C/a$. Substituting $r = ax$ into the two-dimensional relative-motion Hamiltonian gives a dimensionless spectrum that depends only on $q$. The physical-unit potential is $V(r) = -[\pi C/(2r^*)][H_0(r/r^*) - Y_0(r/r^*)]$, where $H_0$ and $Y_0$ are the Struve and Bessel functions.

We compute an 800-q RK table over q=0.25-15 using a cell-centered finite-volume solver. $S_{2p}$ (q) is strictly decreasing and $T_{2p}$ (q) is strictly increasing over this domain.

We solve the radial equation using a cell-centered finite-volume scheme transformed to a symmetric tridiagonal problem. Maximum dimensionless transfer-manifold drift is below $1.45\times10^{-4}$ across the tested grid variants. Calculations are performed with two radial grid setups ($h = 0.09 / r_{max} = 145$ and $h = 0.05 / r_{max} = 220$), with results reported separately in Notes 14 and 15.

**Supplementary Note 2. Same-location source data and uncertainty propagation**

Primary peak centers (eV), in the order 1s/2p/2s/3p, are $MoSe_2$ (1.6390, 1.7797, 1.7983, 1.8182), $WS_2$ (2.0601, 2.1944, 2.2033, 2.2272), and $WSe_2$ (1.6982, 1.8203, 1.8273, 1.8453) (*14, 15*). The source Table 1 explicitly contains $WSe_2$ 3p, drawn from the authors' earlier $WSe_2$ report; it is not merely a derived secondary benchmark. The 2025 correction changes the instrumental resolution from ~6 to ~0.6 meV and clarifies that tabulated errors also reflect fitting uncertainty (*21*).

We perform 200,000 independent Gaussian peak perturbations (seed 20260903) and all 16 sign combinations of the four errors at each of ±1σ, ±2σ and ±3σ. The computed ±2σ ranges are $MoSe_2$ -12.31 to -2.55 meV and $WS_2$ +1.00 to +10.10 meV; at ±3σ, $MoSe_2$ remains negative (-14.80 to -0.17 meV), whereas $WS_2$ spans −1.31 to +12.35 meV. These are full simultaneous deterministic corners, not a probability statement that all 2σ perturbations jointly occur with 95% confidence. The source error combines instrument and fit contributions, and Gaussian independence is an analytic convention, not source-verified covariance. An independent numerical check is summarized in Note 15.

All primary transition energies and their quoted errors are taken directly from the cited source publications (*14, 15*). The source authors have also deposited numerical spectra under DOI 10.57723/289127 (*20*). The primary analysis uses the authors' tabulated peak centers; no optical line-shape refit is performed.

All primary transition energies are taken directly from Refs. (*14, 15*); the numerical spectra are deposited in Ref. (*20*).

**Supplementary Note 3. One-parameter extensions and pure-Coulomb benchmark**

The Fig. 2 extension results are a separate protocol from the same-location three-state primary closure: high-field literature q fixes the scalar RK shape, source-sample 1s–2s fixes scale, and both p states are withheld from scalar-baseline calibration. Three p-specific one-parameter alternatives ($q_p$, a constant p defect, and an $N^{-3}$ localized-core term) are fitted on 2p and tested on 3p. The computed 3p RMSE values are 4.34 meV for scalar RK versus 4.18, 4.10 and 4.19 meV. In the reverse $q_p$ test fitted to 3p then tested on 2p, we find approximately 18 meV versus approximately 2.9 meV for scalar RK in that reciprocal direction. These are different prediction directions; inversion of $q_p$ is nonlinear and need not give equal RMSE. The origin of the unusually large reverse error (conditioning, multiple roots, model rigidity, or physical n-dependence) cannot be established from the diagonal energies alone, so no mechanistic inference is assigned to the asymmetry.

Under the same secondary high-field-q and same-sample 1s–2s scale protocol, held-out 2p/3p RMSE is 11.10 meV for a pure 2D Coulomb reference and 3.68 meV for scalar RK, a 66.8% reduction. The Coulomb reference is intentionally simple and enforces 2s/2p degeneracy; this is a screening-compression comparison, not proof of a unique microscopic baseline. No exact matched pure-Coulomb 1s/2p/2s→3p contrast exists without adding a new calibration prescription, because pure Coulomb cannot fit the observed 2s–2p splitting as its own independent degree of freedom.

**Supplementary Note 4. Spectral compression does not imply state convergence**

This note supplies the scale-separation argument used to interpret the coexistence of strong scalar spectral compression and structured residual information in Fig. 2b. Quantum-defect theory establishes that sufficiently localized departures from an attractive Coulomb tail can be encoded by short-range boundary data or a finite defect (*17, 18*). The narrower point needed here is that convergence of fractional energies does not by itself force convergence of state-sensitive information.

For a two-dimensional Coulomb reference $H_C = -(1/2)\nabla^2 - 1/r$, the standard bound-state convention used here gives $E_n^C = -2/(2n-1)^2$. Writing $N = 2n-1$, the binding energy scales as $B_n^C = |E_n^C| \propto N^{-2}$. For a fixed level offset $k$ and $n \gg 1$, $E_{n+k}^C = -2/(N+2k)^2$, and a Taylor expansion gives $E_{n+k}^C - E_n^C = 8k\,|\,N^{-3} + O(N^{-4})\,|$; adjacent high-Rydberg spacings therefore scale as $n^{-3}$.

For a sufficiently localized $V_{\text{core}}$ and a penetrative fixed-angular-momentum channel with a finite threshold, the normalized two-dimensional Coulomb radial functions carry an $N^{-3/2}$ leading high-$n$ normalization factor at fixed core radius. This follows directly from the normalized 2D Coulomb functions: the $q_0^{3/2}$ prefactor with $q_0 \propto N^{-1}$ supplies the leading fixed-core normalization, while the fixed-angular-momentum polynomial factors have a finite threshold. The product of two adjacent high-$n$ core amplitudes therefore gives $\langle n | V_{\text{core}} | n+k \rangle \propto N^{-3}$, because the radial integration is confined to a dimensionless core region. The diagonal correction has the same scale, $\delta E_n \propto N^{-3}$. Two dimensionless limits then differ: $|\delta E_n| / B_n^C \propto N^{-1} \to 0$, whereas $|\langle n | V_{\text{core}} | n+k \rangle| / |E_{n+k}^C - E_n^C|$ can remain $O(1)$.

This hierarchy explains why the 66.8% reduction of the pure-Coulomb error baseline and the persistence of material-selective p-channel residuals are not contradictory: total binding energy or aggregate spectral error is not the only relevant scale for state-resolved information. It also clarifies that the Fig. 2 result is complementary rather than redundant. In that particular scalar-versus-p-specific static contrast, the observable model separation itself decreases from 3p to 6p, so higher $n$ supplies less experimental discrimination even though spectral convergence alone would not guarantee Hilbert-space state convergence.

The applicability boundary is essential. The $n^{-3}$ matrix-element statement above is a Coulomb-tail, localized-core scale argument for a penetrative fixed-angular-momentum sector; centrifugal suppression, nonlocal band-spin structure, Berry/geometric terms, extended or dynamic screening, interlayer dressing, phonons, and continua can alter the relevant matrix elements and denominators. The measured 2p/3p states are also low members of the ladder rather than an asymptotic sequence. We therefore use this argument as a physical counterexample to the inference “small spectral error implies negligible state information,” not as a mechanism assignment for the observed p-state residuals.

**Supplementary Note 5. Independent observables and data provenance**

The literature controls are separated by information flow. Ref. (*14*) provides the published primary same-location s/p energies. Ref. (*16*) supplies an independent-platform $WS_2$ directional stress, ref. (*7*) provides size controls, ref. (*24*) a separate optical comparison, ref. (*9*) a tunable dielectric-screening axis, and ref. (*25*) independent dark-exciton context.

For the auxiliary $MoSe_2$ global-size check, we digitized only the independently published B=0 1s/2s/3s peak-center positions from the relevant figure in ref. (*7*); no public numeric trace or table for precisely those centers was available. We excluded source-derived masses, screening lengths, calculated energies and theoretical radii from the independent input because the original source fit drew on more than the zero-field peaks. The previously reported energy-only inversion predicts an rms-radius-squared ratio of approximately 4.59 for 3s/2s, compared with a value 6.41 from reported radii of 8.1 and 3.2 nm; ±2-meV digitization perturbations give approximately 4.32–4.88. This is an auxiliary stress test, not a raw-spectrum fit, an independent experiment or evidence for a unique mechanism. The $WSe_2$ energy-to-diamagnetic-size comparison is likewise secondary (*2, 8*).

**Supplementary Note 6. Fixed-q curvature and parameter provenance**

The external-*q* representation is deliberately placed downstream of the parameter-free same-location 3p result. High-field RK parameter sets (*2, 6*) yield $q = 2.665$ for $MoSe_2$, $q = 1.188$ for $WS_2$, and $q = 1.680$ for $WSe_2$. These model-derived coordinates are held fixed; numerical pole locations and the $Q_{4p}$ interval are therefore conditional on fixed *q*. The numerical inputs are listed in Table S1.

At a specified fixed external RK shape q, corrections of the form $\Delta E = A + B{\cdot}E^0$ are removed exactly by $\mathrm{Q_p} = \left(\Delta E_\mathrm{p} - \Delta E_\mathrm{1S}\right) - r_\mathrm{p}\left(\Delta E_\mathrm{2S} - \Delta E_\mathrm{1S}\right)$, with $r_\mathrm{p} = \left(E^0{}_\mathrm{p} - E^0{}_\mathrm{1S}\right)/\left(E^0{}_\mathrm{2S} - E^0{}_\mathrm{1S}\right)$. The divided difference $C_\mathrm{12;p} = \mathrm{Q_p}/\left[\left(E^0{}_\mathrm{p} - E^0{}_\mathrm{1S}\right)\left(E^0{}_\mathrm{p} - E^0{}_\mathrm{2S}\right)\right]$ measures residual curvature in that fixed reference. This is affine invariance, not elimination of arbitrary n-dependent screening errors, q mismatch or cross-sample structure; quadratic and other nonaffine baseline corrections survive and could change the material fingerprints. Independent kernel and q-variation sensitivity are needed for mechanism attribution.

**Supplementary Note 7. Minimal single-pole dark-state model**

The tested representation assumes (A1) a fixed external-q RK P-space baseline, (A2) a passive Hermitian dark-state sector, (A3) a common scalar positive spectral measure sampled by p states, (A4) a projected/linearized diagonal self-energy for the curvature inversion, and (A5) an external q held fixed without cross-device probability propagation. The resulting one-pole admissibility is a model-dependent mathematical test; a failure of the ansatz is not a proof that the material necessarily has several microscopic dark states. The source data do not measure actual pole locations or a unique matrix spectral rank.

**Supplementary Note 7A. Exploratory branch identity and higher-pole identifiability**

The fixed-q quotient results and Fig. 3 classify the projected scalar common-residue one-pole construction described above. They should not be promoted into universal theorems about every possible passive Hermitian Hamiltonian. We therefore performed a separate exploratory self-consistent finite-pole stress test using the same published center energies, the external-q values used in the main analysis ($MoSe_2$ 2.665, $WS_2$ 1.188, $WSe_2$ 1.680), and the same-sample 1s–2s scale. No unmeasured 4p energy is used for fitting.

For a selected RK-like bare level $x_i$ and observed level $y_i$, write the finite-pole eigenvalue equation as $y_i = x_i + A + Bx_i + \Sigma_j\, g_j^2 / \left(y_i - \omega_j\right)$, with $g_j^2 \geq 0$. A and B are eliminated using the 1s/2s anchors. For each trial set of pole locations, the measured 2p and 3p centers constrain the positive residues. Diagonalizing the corresponding real symmetric Hamiltonian gives each candidate eigenbranch and its RK-like projector weight Z. The supplementary branch-continuity criterion $Z > 0.5$ is an illustrative modeling choice, not a source-measured transition intensity or a universal physical threshold.

$MoSe_2$: the center-data single-pole solution of this particular self-consistent stress has $g^2 < 0$ and is excluded by positive weight. Yet positive-weight higher-pole counterexamples can reproduce all four existing peaks and yield both signs of $Q_{4p}$; one four-pole witness has $Q_{4p} = 0$, $E_{4p} =$ 1.838282 eV. These poles are mathematical constructs, not experimentally observed dark excitons. Thus the present diagonal energies do not support a narrow model-independent 4p forbidden band.

$WSe_2$: a self-consistent positive-residue scalar one-pole solution can reproduce the central energies but places the observed 2p branch at $Z_{2p} \approx 0.275$, predominantly dark-state-like under that selected basis. This is different from the projected common-residue compatibility test in the main text. Positive-weight three-pole witnesses also exist for $WSe_2$ with $Q_4$p values differing from the one-pole continuation, further illustrating model non-identifiability rather than establishing competing experimentally supported predictions. A meaningful measurement must constrain transition matrix elements and phases in addition to peak positions.

Independent dark-exciton ns and B:1s measurements (*14, 26–28*) motivate additional experiments but do not independently locate every p-symmetry pole that can couple to the measured bright A p series. In particular, an observed B:1s near the A:4p spectral region does not establish symmetry-allowed direct p-channel coupling. Neither its cross-sample absolute energy nor a dark 1s splitting is substituted for an unmeasured dark-p pole. Only after the allowed pole support, selection rules, and coupling bounds are measured independently can one freeze a material-specific 4p exclusion range.

Under A1-A5, a one-pole representation gives $C_p = g^2 / [(E_1 - \omega)(E_2 - \omega)(E_p - \omega)]$ . The curvature ratio fixes $\omega$, after which the inferred residue must be nonnegative for a passive Hermitian realization. $WS_2$ satisfies this condition in the one-pole common-measure model. $MoSe_2$ would require a negative one-pole residue, and $WSe_2$ does not match the tested common-residue one-pole form. Diagonal energies do not determine the matrix spectral rank or residue vectors.

**Supplementary Note 8. Scope of microscopic sensitivity analyses**

The microscopic stress tests are selected low-complexity sectors. Independently parameterized Berry-curvature and quantum-metric terms, quasi-degenerate trigonal p-d mixing, lattice-local derivative-contact corrections, finite orbital-exchange bounds, and smooth long-wavelength screening completions are all evaluated without fitting the target 3p residual. Their purpose is to bound simple intrinsic/static patches and to define the observables needed for further discrimination. Sharp finite-$q$ screening, intervalley/local-field structure, nonlocal spin mixing, interlayer dressing, phonons, and continua remain possible microscopic realizations.

The two-band Berry curvature / effective mass sensitivity range used in the global robustness matrix spans a range of Berry/mass parameter variations; the sign conclusions are robust across the envelope. A Mo/W conduction-band spin-splitting sign reversal does not by itself reverse the A-bright joint exciton Berry-curvature sign at a fixed valley in the independently parameterized three-band calculation.

**Supplementary Note 9. Three-band Berry curvature calculation and the 15.9 meV Berry-curvature-induced 2p splitting**

The three-band calculation uses the nearest-neighbor GGA parameters of (*23*) together with the metal on-site spin-orbit coupling and the tabulated conduction-band spin splitting. Because the nearest-neighbor three-band Hamiltonian leaves the conduction-edge spin splitting incomplete, the tabulated first-principles K-point conduction-band splitting is imposed explicitly as a spin-dependent diagonal shift of the $d_{z^2}$ orbital. The numerical band inputs are listed in Table S2.

At K, the A-bright $MoSe_2$ joint exciton Berry curvature is $\Omega_{\mathrm{exc}}$=-31.7092818 $Å^2$. Separately, the scalar physical-unit RK calibration uses only 1s/2p/2s with $\mu = 0.350 m_0$ and gives $\kappa$ =4.229376275 and $r_0$=4.548148296nm. The resulting 2p wavefunction gives $A_2 = \langle V'(r)/r \rangle_{2p}$ =0.05014131154 $\mathrm{eV\,nm^{-2}}$. Since $|\Omega_{\mathrm{exc}}|$=0.3170928179 $\mathrm{nm}^2$, the Berry-curvature-induced splitting between the $m = \pm 1$ branches is $|\Omega_{\mathrm{exc}}| A_2 = 0.01589944977$ eV = 15.89944977 meV. The Darwin term is identical for the two $m = \pm 1$ branches and cancels in the separation.

**Supplementary Note 10. Observability of optical phase and branch weights**

In coherent spectroscopy, two nearby resonances with different complex amplitudes can produce similar intensity profiles even when their relative phases and branch weights differ. The elementary two-pole construction illustrates why independent phase or symmetry information would help distinguish possible microscopic branches. The present study does not fit new optical spectra or infer experimental branch weights from the published intensity traces. Model-level observability is discussed here to motivate independent future measurements, while the reported experimental peak positions and errors remain those of the original source authors.

**Supplementary Note 11. Robustness checks**

The analysis distinguishes source-reported peak centers and errors, deterministic peak-error corners and Gaussian propagation, RK transfer and leave-one-out checks, cross-device high-field-q comparisons, minimal static p-sector extensions, selected band-geometry sensitivity estimates, grid variations, and exploratory positive-weight counterexamples.

**Supplementary Note 12. Independent-platform $WS_2$ sign consistency check**

Shi et al. (*16*) give the separate room-temperature plasmon-enhanced $WS_2$ energies A(1s)=2.00, A(2p)=2.21, A(2s)=2.24, A(3p)=2.35 eV. Applying the same 1s/2p/2s → 3p extrapolation protocol gives a predicted 3p≈2.28286 eV and a residual of +67.14 meV. This is roughly twelve times the hBN $WS_2$ central residual of +5.58 meV. We include it as a qualitative same-sign stress test across very different temperature, screening, plasmonic architecture, line extraction and covariance conditions.

**Supplementary Note 13. $WS_2$ 4p prediction and uncertainty propagation**

For the stated externally fixed-q, positive common-scalar one-pole representation, $WS_2$ 2p and 3p quotient curvatures determine the pole and residue and yield a frozen next-state prediction $Q_{4p}$=+4.08 meV. The plotted +4.1 meV is rounding, not an updated fit. The already available 3p input helped choose and constrain this model; its prospective status rests only on the future same-platform 4p datum. A different admissible model class could yield a different 4p outcome.

The supplied fixed-q numerical propagation yields a median of approximately +4.0 meV and a conditional 68% interval of +2.9 to +5.0 meV. This interval assumes the one-pole model and does not include cross-device q uncertainty or future experimental covariance; it is not a 95% falsification threshold. A meaningful prospective score should freeze the point target, specify independent measurement preprocessing, incorporate the observed covariance and independently constrained q, and report the selected model family before accessing the new 4p data.

All calculations were performed in MATLAB using a cell-centered finite-volume RK solver. The same solver is used for the RK transfer manifold, leave-one-out sensitivity, external-q comparison, deterministic error boxes, and Gaussian uncertainty propagation. The experimental source Table 1 and public numerical spectra have their own published DOI (10.57723/289127) (*20*).

**Supplementary Note 14. Alternative held-out states and external-q consistency**

A cell-centered RK solver (h=0.05 in dimensionless radius, radial domain 220) computes the same-location 1s, 2p, 2s, 3p spectrum directly. For each of the four leave-one-level-out configurations we solve q from the three calibration levels, then determine the offset and common scale from two of them before evaluating the excluded fourth state. q is constrained to the originally declared interval 0.25–15. This is a retrospective training-set sensitivity audit, not four independent cross-validation observations and not an unbiased multiple-testing corrected significance test. Its results confirm that the sign reversal describes the physically motivated p-to-p transfer protocol, rather than an invariant feature of every possible triplet. Mesh refinement from $h = 0.09 / r_{max} = 145$ to $h = 0.05 / r_{max} = 220$ changes the central 3p residual by only ~0.02–0.03 meV.

Table S4. Independent leave-one-level-out sensitivity for three materials. Residual= measured−predicted; all values are approximate meV, computed with the finite-volume RK implementation.

| Material | Excluded | Calibration triplet | Residual (meV) | q_fit |
|---|---|---|---|---|
| $MoSe_2$ | 1s | 2p/2s/3p | no q root in 0.25–15 | — |
| $MoSe_2$ | 2p | 1s/2s/3p | −9.38 | 0.769 |
| $MoSe_2$ | 2s | 1s/2p/3p | +3.86 | 1.971 |
| $MoSe_2$ | 3p | 1s/2p/2s | −7.39 | 3.361 |
| $WS_2$ | 1s | 2p/2s/3p | +66.67 | 0.478 |
| $WS_2$ | 2p | 1s/2s/3p | +7.09 | 2.966 |
| $WS_2$ | 2s | 1s/2p/3p | −3.05 | 1.432 |
| $WS_2$ | 3p | 1s/2p/2s | +5.57 | 0.865 |
| $WSe_2$ | 1s | 2p/2s/3p | +26.44 | 0.521 |
| $WSe_2$ | 2p | 1s/2s/3p | +3.01 | 1.292 |
| $WSe_2$ | 2s | 1s/2p/3p | −1.23 | 0.888 |
| $WSe_2$ | 3p | 1s/2p/2s | +2.21 | 0.692 |

In the primary triplet, the same-location inferred q differs from the high-field literature values (Table S5); in $WSe_2$ this mismatch even changes the residual direction when the source-sample 1s–2s scale is combined with fixed external q. This comparison is informative about reference dependence, not a clean high-field/low-field validation, because q derives from different RK analyses, devices, environmental conditions and independent source assumptions. Importantly, $q = r^* / a = 2\mu C r^* / \hbar^2$ scales approximately as $\mu r_0/\kappa^2$ for the stated conventions: larger q is not universally 'weaker screening'.

Table S5. Same-location q from the primary 1s/2p/2s triplet versus literature high-field model-derived q. The last column is a secondary fixed-q/1s–2s 3p residual, not the main closure.

| Material | same-location q | external q | fixed-q 3p residual (meV) |
|---|---|---|---|
| $MoSe_2$ | 3.361 | 2.665 | −6.09 |
| $WS_2$ | 0.865 | 1.188 | +4.29 |
| $WSe_2$ | 0.692 | 1.680 | −1.06 |

**Supplementary Note 15. Deterministic error boxes, solver comparison, and numerical consistency**

We performed a distinct 16-corner recomputation for each material at ±1σ, ±2σ and ±3σ using the source Table 1 peak errors, with the finite-volume RK solver described in Note 14. This finite-volume calculation reproduces the core sign patterns: $MoSe_2$ and $WS_2$ have opposite signs across the full ±2σ range; $WS_2$, but not $MoSe_2$, crosses zero in extreme ±3σ combinations. $WSe_2$ already crosses zero at some ±1σ deterministic corners. The corrected source errors are peak-position errors combining spectral resolution and fitting error; the error boxes are not a joint 95% or 99.7% confidence region.

Table S6. Independent peak-error-corner recomputation; values in meV. These values use the finite-volume solver; Fig. 1 uses precomputed manifold interpolation.

| Material | Box | Min | Max | Sign fixed |
|---|---|---|---|---|
| $MoSe_2$ | ±1σ | -9.74 | -5.06 | yes |
| $MoSe_2$ | ±2σ | -12.11 | -2.74 | yes |
| $MoSe_2$ | ±3σ | -14.50 | -0.44 | yes |
| $WS_2$ | ±1σ | +3.32 | +7.72 | yes |
| $WS_2$ | ±2σ | +1.07 | +9.94 | yes |
| $WS_2$ | ±3σ | -1.24 | +12.22 | no |
| $WSe_2$ | ±1σ | -0.08 | +4.33 | no |
| $WSe_2$ | ±2σ | -2.30 | +6.62 | no |
| $WSe_2$ | ±3σ | -4.56 | +8.97 | no |

The finite-volume solver's $MoSe_2$ ±2σ limits (-12.11, -2.74 meV) differ by approximately 0.2 meV from the manifold-interpolated bounds (-12.31, -2.55 meV); both agree on sign and all qualitative conclusions, and the $WS_2$ limits agree within ~0.01 meV. The difference reflects the finite-volume solver versus precomputed manifold interpolation. An 800-q table confirms monotonic $S_{2p}$ and $T_{3p}$ over 0.25–15.

**References**

References are numbered identically to the main manuscript; no separate supplementary-only bibliography is used. All citations in the Supplementary Materials refer to the unified reference list (Refs. *1–30*) in the main manuscript. Refs. *24–30* are cited only in the Supplementary Materials.

**Fig. S1.**

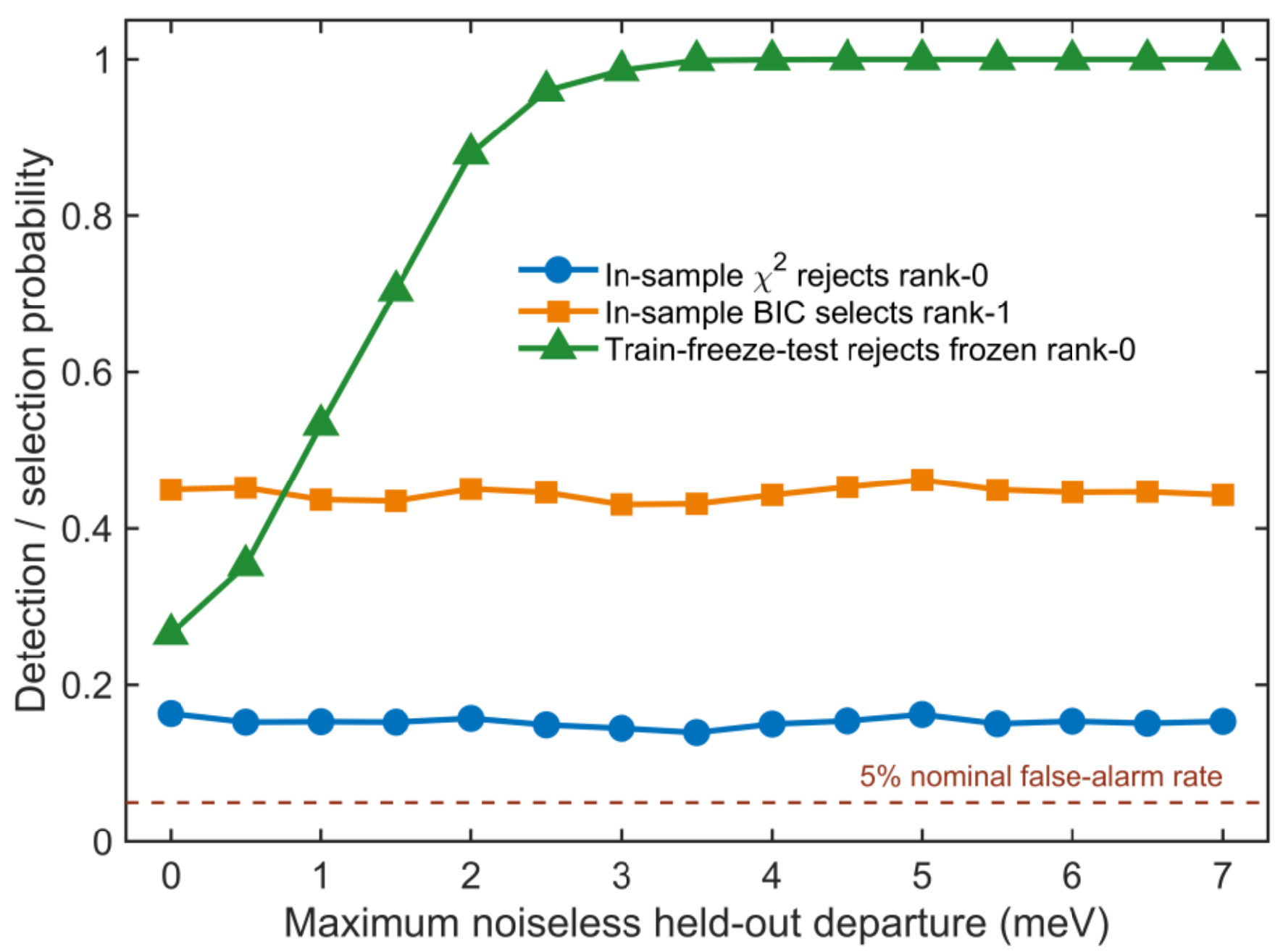

**Fig. S1. Frozen synthetic transfer-power benchmark.** Known-ground-truth synthetic spectra calibrate the distinction between training fit and held-out transfer. At the declared 0.5-meV noise / 2.5-meV held-out-departure setting, in-sample diagnostics have low rejection power while the predeclared held-out test is substantially more sensitive. Synthetic benchmarks calibrate the analysis logic.

**Fig. S2.**

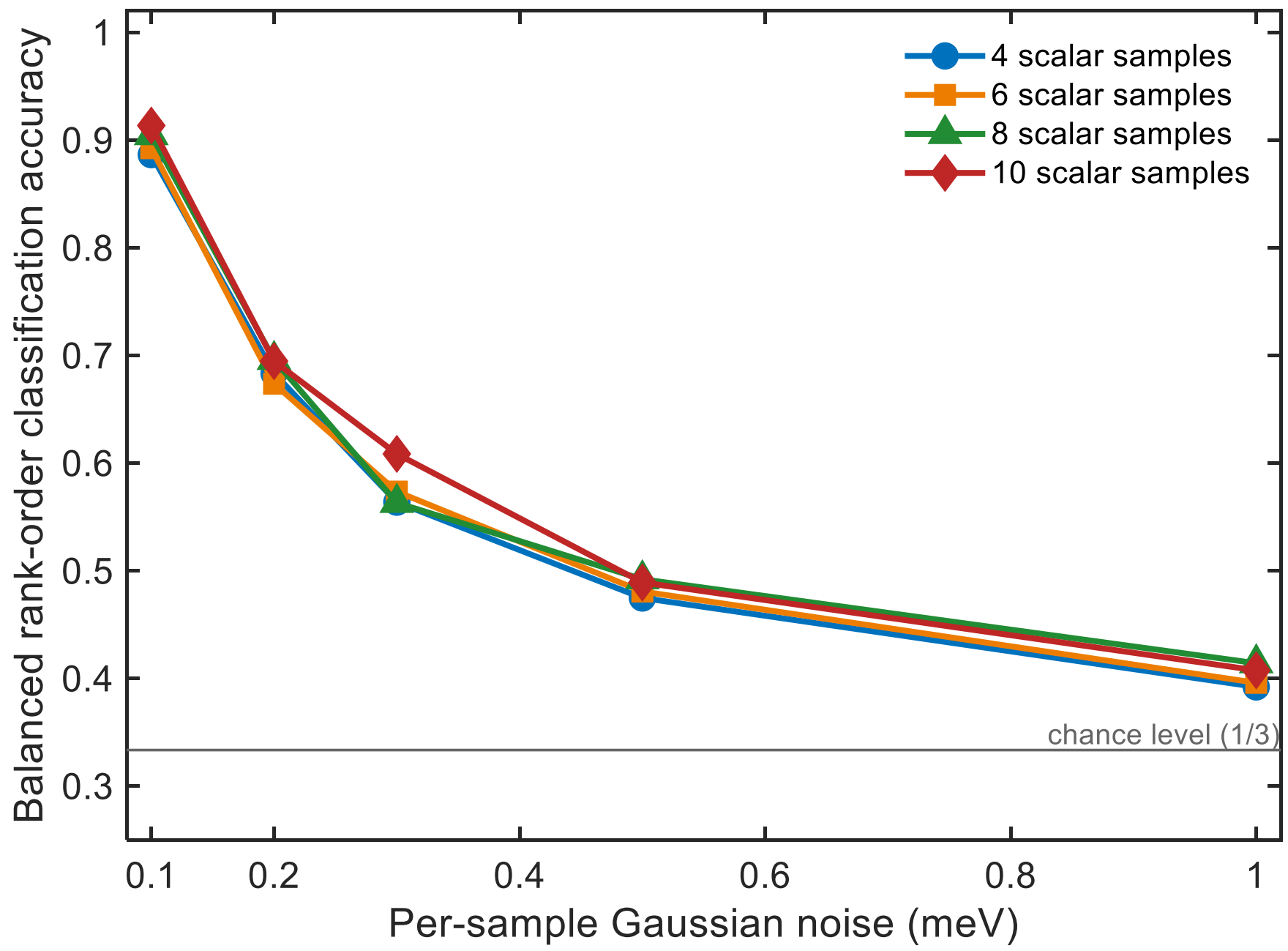

**Fig. S2. Precision controls realization-order recovery in the synthetic generator.** Increasing the number of low-precision diagonal samples provides little benefit in the tested generator, whereas higher precision markedly improves recovery. "Order" denotes scalar/common-measure finite-pole realization order.

**Fig. S3.**

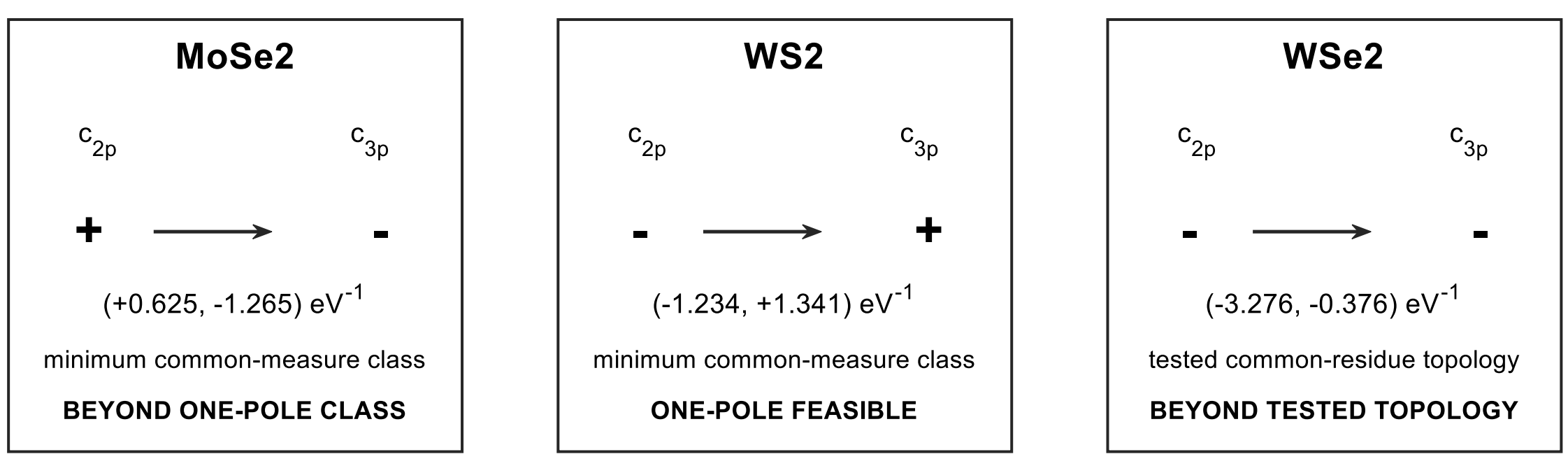

**Fig. S3. Residual orbital structure as a material fingerprint: classification under minimal dark-state models.** The three panels show $MoSe_2$, $WS_2$, and $WSe_2$, respectively, with the 2p-to-3p sign pattern and curvature values ($eV^{-1}$) for each material. The projected curvature values shown in Fig. 3 are $MoSe_2$ (+0.625, -1.265), $WS_2$ (-1.234, +1.341), and $WSe_2$ (-3.276, -0.376). Under the tested passive one-pole/common-residue construction, $MoSe_2$ lies beyond the one-pole class because it would require a negative residue, $WS_2$ is one-pole feasible, and $WSe_2$ lies beyond the tested common-residue topology.

**Fig. S4.**

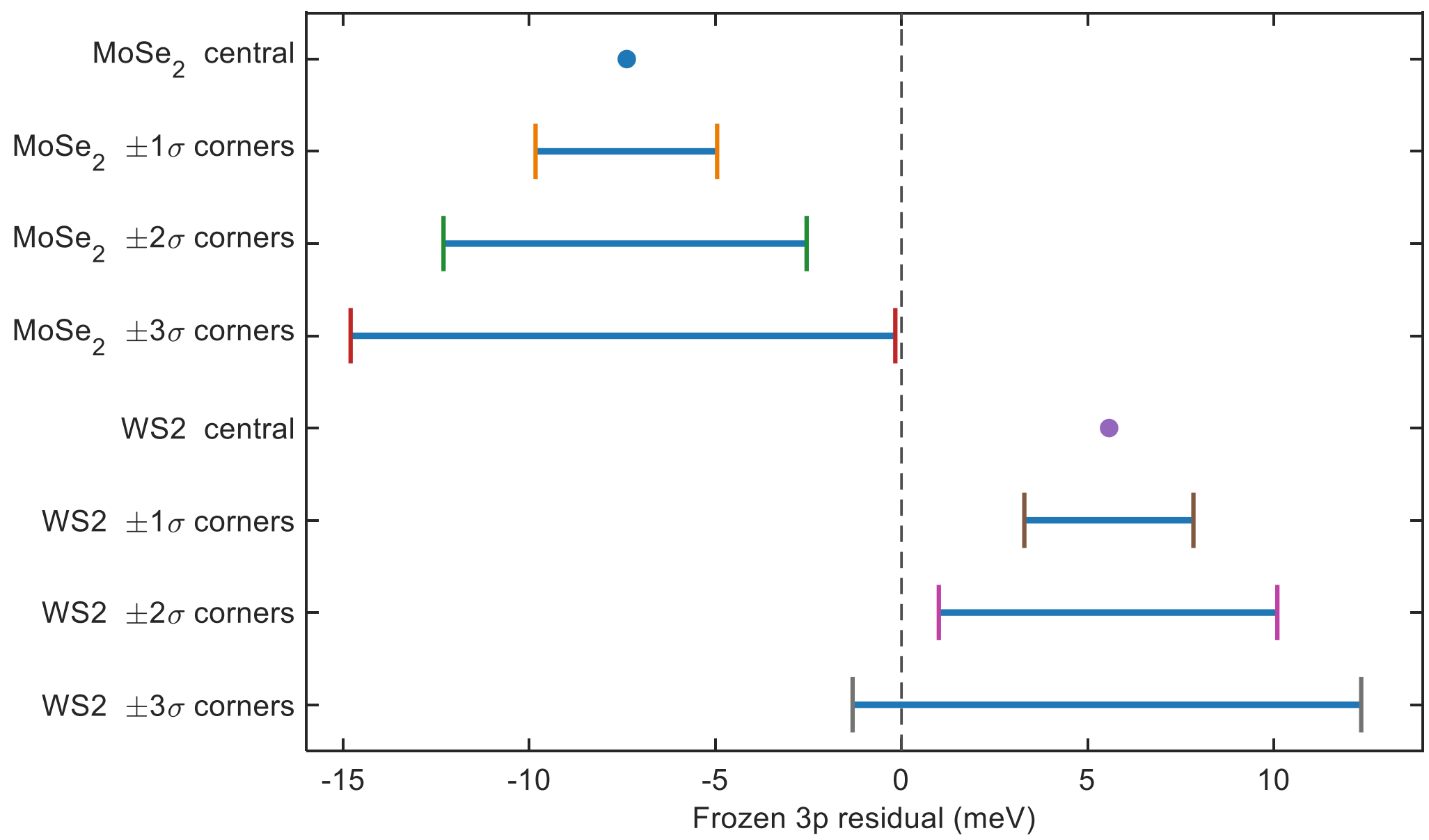

**Fig. S4. Complete deterministic error-box evaluation.** The $MoSe_2$/$WS_2$ sign pair is unchanged at every simultaneous $\pm 1\sigma$ and $\pm 2\sigma$ corner. $MoSe_2$ remains negative at all $\pm 3\sigma$ corners; $WS_2$ crosses zero only in the extreme $\pm 3\sigma$ box. Central values are shown as dots ($MoSe_2$=-7.38meV, $WS_2$=+5.58meV), and the horizontal bars span the full 16-corner deterministic range at each $\sigma$ level.

**Fig. S5.**

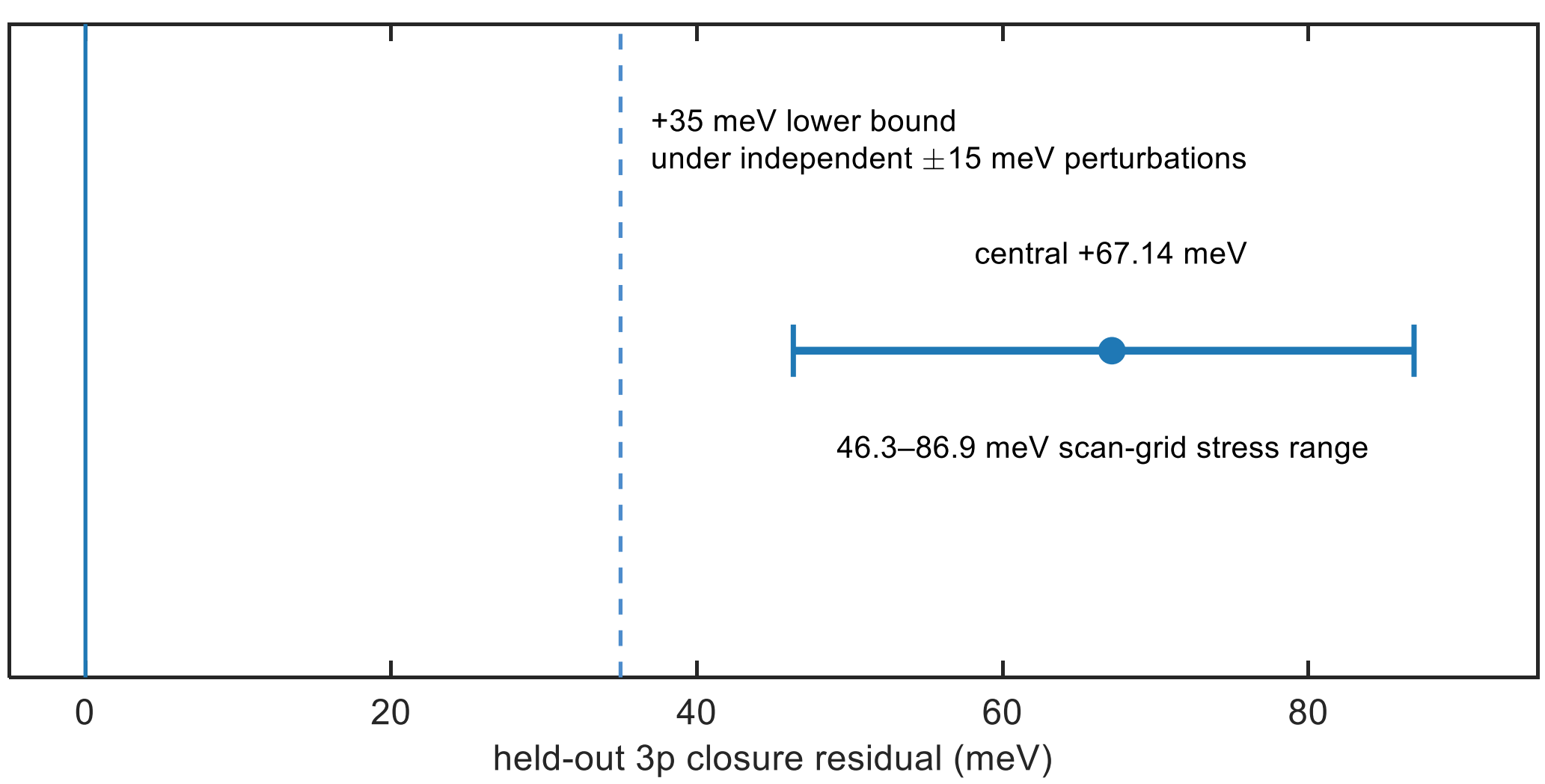

**Fig. S5. Independent-platform $WS_2$ sign consistency test.** The room-temperature plasmon-enhanced data of (*16*) give a positive 3p residual under broad deterministic perturbations. The central +67.14-meV value is shown with the 46.3 to 86.9 meV scan-grid stress range; the +35 meV dashed line marks the lower bound under independent ±15 meV perturbations of all four levels. The central value is kept out of the main Fig. 1 comparison because cross-platform magnitudes are not quantitatively pooled.

**Table S1.**

| **Material** | $r_0$ (nm) | $\kappa$ | $\mu/m_0$ | $q$ | **Source/role** |
|---|---|---|---|---|---|
| $MoSe_2$ | 3.90 | 4.40 | 0.350 | 2.665 | Ref.(*6*); fixed-$q$ analysis |
| $WS_2$ | 3.40 | 4.35 | 0.175 | 1.188 | Ref. (*6*); fixed-$q$ analysis |
| $WSe_2$ | 4.50 | 4.50 | 0.200 | 1.680 | Ref. (*2*); fixed-$q$ analysis |

**Table S1. External high-field RK coordinates used only in the fixed-q analysis.**

**Table S2.**

| **Parameter** | $MoSe_2$ | $WS_2$ | **Units/source role** |
|---|---|---|---|
| $a$ | 3.326 | 3.191 | Å; three-band lattice constant |
| $\varepsilon_1$ | 0.919 | 1.130 | eV |
| $\varepsilon_2$ | 2.065 | 2.275 | eV |
| $t_0$ | -0.188 | -0.206 | eV |
| $t_1$ | 0.317 | 0.567 | eV |
| $t_2$ | 0.456 | 0.536 | eV |
| $t_{11}$ | 0.211 | 0.286 | eV |
| $t_{12}$ | 0.290 | 0.384 | eV |
| $t_{22}$ | 0.130 | -0.061 | eV |
| $\lambda$ | 0.091 | 0.211 | eV |
| $\Delta E_c$ | -0.021 | +0.029 | eV; conduction-band spin-splitting input |

**Table S2. Three-band and spin-splitting inputs used in the Berry-curvature sensitivity calculation (*23, 29, 30*).**

**Table S3.**

| **Quantity** | **Value** | **Units** |
|---|---|---|
| $\Omega_{exc}$ | -31.7092818 | $eV/nm^2$ |
| $\lvert\Omega_{exc}\rvert$ | 0.317092818 | $nm^2$ |
| $\kappa$ | 4.22937627 | dimensionless |
| $r_0$ | 4.54814830 | nm |
| $A_2 = \langle V' / r \rangle_{2p}$ | 0.05014131154 | $eVnm^2$ |
| $\Delta_{2p}$ | 15.89944977 | meV |

**Table S3. Derived quantities from the Berry-curvature sensitivity calculation.**